\documentstyle[epsfig]{elsart}
\begin{document}

\begin{frontmatter}
\title{High-momentum tail in the Tonks gas under harmonic confinement}
\author{A. Minguzzi, P. Vignolo and M. P. Tosi}
\address{NEST-INFM and Classe di Scienze,
Scuola Normale Superiore,
Piazza dei Cavalieri 7, I-56126 Pisa, Italy}

\maketitle

\begin{abstract}
We use boson-fermion mapping to show that the single-particle momentum 
distribution in a one-dimensional gas of hard point-like bosons 
(Tonks gas) inside a harmonic trap decays as $p^{-4}$ at large momentum $p$. 
The relevant integrals expressing the one-body density matrix are 
evaluated for small numbers of particles in a simple Monte Carlo 
approach to test the extent of the asymptotic law and to illustrate 
the slow decay of correlations between the matter-wave field at 
different points.
\end{abstract}

\end{frontmatter}    

\section{Introduction}
The realization of Bose-Einstein condensates of lower 
dimensionality in suitable optical and magnetic traps \cite{art1,art2}, and 
advances in atom waveguide technology with potential applications to 
atom interferometry and integrated atom optics 
\cite{art3,art4,art5,art6,art7}, give special 
motivation to theoretical studies of dilute atomic fluids in a 
regime where the quantum dynamics becomes quasi-one-dimensional (1D). 
The condensate fraction is depleted in this situation even at zero 
temperature, and the standard stationary or time-dependent 
Gross-Pitaevskii theory loses its usefulness \cite{art8,art9,art10}. 
An extreme 
limiting example is provided by the so-called Tonks gas of hard 
point-like bosons, in which the transverse confinement is taken 
to be so strong that the dynamics reduces to strictly 1D motion. 
The occupation of the lowest orbital behaves in this case with the 
number $N$ of particles as $N^n$ with $n < 1$~\cite{art11}. 
The momentum distribution 
still has a strong peak on approaching zero momentum \cite{art8,art11}, and 
coherence effects have been shown to be present in the form of Talbot 
recurrences following an optical lattice pulse~\cite{art12} and of dark 
soliton-like behaviour in response to a phase-imprinting pulse~\cite{art13}.

A number of properties of the Tonks gas of impenetrable bosons 
can be calculated exactly through a mapping into an ideal 1D gas of 
spin-polarized fermions~\cite{art14}. In both fluids each particle behaves as 
a hard point which is at any time free to move on a segment comprised 
by its two immediate neighbours and can exchange with either of them on 
contact. Considering in particular the many-body ground state, the 
boson-fermion mapping consists of the identity between the bosonic wave 
function and the modulus of the fermionic wave function~\cite{art14}. This of 
course implies that, whereas all configurational probability distributions 
are the same in the two fluids, their single-particle momentum 
distributions are in general completely different~\cite{art15}.

A visualization of the ground state can be obtained by varying the 
position of one particle while all others are kept fixed. The fermionic 
wave function varies smoothly from positive to negative across each node 
so as to mimimize the kinetic energy, and hence the bosonic wave function, 
being everywhere non-negative, has a cusp at each node. We show in this 
paper that from this cusp condition the momentum distribution of the 
Tonks gas under harmonic confinement acquires a long tail at high 
momentum $p$, decaying asymptotically as $p^{-4}$. We also gauge the extent 
of this asymptotic behaviour by Monte Carlo calculations of the momentum 
distribution for the case of two particles in a harmonic well, where 
high statistical accuracy and simple analytical results are easily obtained.

\section{Asymptotic tails of the bosonic momentum distribution}
The single-particle momentum distribution $n(p)$ at momentum $p=\hbar k$
is the Fourier transform of the one-body density matrix
$\rho(x,x')$ with respect to the relative coordinate $r=x-x'$, averaged
over the centre-of-mass coordinate $R=(x+x')/2$.
The ground-state wave function for a Tonks gas of $N$ bosons under 
longitudinal harmonic confinement has been calculated by Girardeau
{\it et al.}~\cite{art11}, with the result 
\begin{equation}
\psi_0^B(x_1,\dots,x_N)=C_N\prod_{1\le j<k\le N}|x_k-x_j|
\prod_{i=1}^N\exp(-x_j^2/2x^2_{ho}).
\end{equation}
Here, $C_N$ is a normalization constant and $x_{ho}=(\hbar/m\omega)^{1/2}$
is the harmonic oscillator length determined by the trap frequency
$\omega$ and the particle mass $m$. Taking $x_{ho}$ as the unit of length,
the momentum distribution is then given by
\begin{eqnarray}
&n(p)&={\cal N}_N\int_{-\infty}^{+\infty}dR\int_{-\infty}^{+\infty}
dr\,\exp[-ikr-(R^2+r^2/4)]
\int_{-\infty}^{+\infty}\!\!dQ_2\dots\int_{-\infty}^{+\infty}
\!\!dQ_N\nonumber\\
&&\times \,\exp\left(-\sum_{l=2}^N Q_l^2\right)
\left\{\left[\prod_{2\le j<k\le N}(Q_k-Q_j)^2\right]
\prod_{i=2}^N\left|(Q_i-R)^2-r^2/4\right|\right\}
\label{eq2}
\end{eqnarray}
where ${\cal N}_N$ is a factor depending on the number of particles.

Let us first treat the case $N = 2$. The momentum distribution is given by
\begin{equation}
n(p)={\cal N}_2\int_{-\infty}^{+\infty}dR
\int_{-\infty}^{+\infty}dQ_2\exp(-R^2-Q_2^2)I(Q_2,R;k)
\end{equation}
where
\begin{equation}
I(Q_2,R;k)=\int_{-\infty}^{+\infty} dr\,|(Q_2-R)^2-r^2/4|
\exp(-ikr-r^2/4).
\end{equation}
The integrand has two singularities of the type $|r\pm\alpha|$ and we 
can use a theorem given in the book of Lighthill [16] to evaluate its 
asymptotic value in the limit $k\rightarrow\infty$. 
Using the result $\int dz|z|\exp(ikz)=-2/k^2$ we get
\begin{equation}
\lim_{k\rightarrow\infty}I(Q_2,R;k)=
-\frac{2|Q_2-R|}{k^2}\cos[2k(Q_2-R)]\exp[-(Q_2-R)^2].
\end{equation}
Introducing the notations $X = (Q_2 - R)/\sqrt{2}$ and 
$Y =  (Q_2 + R)/\sqrt{2}$, the asymptotic behaviour of 
the momentum distribution is then given by
\begin{equation}
\lim_{p\rightarrow\infty}n(p)=\frac{-2\sqrt{2}}{k^2}
{\cal N}_2 \int_{-\infty}^{+\infty}dY\exp(-Y^2)
\int_{-\infty}^{+\infty}dX|X|\cos(2\sqrt{2}kX)\exp(-3X^2)
\end{equation}
and Lighthill's theorem can again be employed to evaluate 
the integral over $X$. The final result is
\begin{equation}
\lim_{p\rightarrow\infty}n(p)=2\sqrt{\frac{2}{\pi}}\frac{(\hbar m\omega)
^{-3/2}}{p^4}\,,
\label{eq7}
\end{equation}
having used the value of the normalization constant $C_2$ reported by 
Girardeau {\it et al.}~\cite{art11}.

In applying the above procedure to the calculation of the asymptotic 
value of Eq. (2) for arbitrary $N$, we notice that the same pair of 
singularities arises for each value of the index $i$ and that they all 
contribute equally to the final result. Choosing $i = 2$, we thus find
\begin{equation}
\lim_{p\rightarrow\infty}n(p)=\frac{A_N}{p^4}\,
\end{equation}
where the coefficient $A_N$ is determined by a multiple integral,
\begin{eqnarray}
&A_N&\propto -(N-1)\,{\cal N}_N \int_{-\infty}^{+\infty}
dY\exp(-Y^2)\int_{-\infty}^{+\infty}
dQ_3\dots\int_{-\infty}^{+\infty}
dQ_N  \,\exp\left(-\sum_{l=3}^N Q_l^2\right)\nonumber\\
&&\times 
\left[\prod_{3\le j<k\le N}(Q_k-Q_j)^2
\left(Q_k-Y/\sqrt{2}\right)\left(Q_j-Y/\sqrt{2}\right)\right].
\end{eqnarray}
We have thus shown that the momentum distribution of the 
Tonks gas in a harmonic trap decays asymptotically as the inverse 
fourth power of the momentum.

\section{Numerical results by a Monte Carlo method}
In this section we evaluate the one-body density matrix and the 
momentum distribution function for a small number of particles by a 
simple Monte Carlo method. The density matrix involves the integrals 
over $(Q_2,\dots,Q_N)$ entering the second line in Eq.~(\ref{eq2}). 
These are calculated by (i) 
extracting for each variable $Q_i$ a random number distributed 
along a Gaussian of unitary width, and (ii) averaging over all trials 
the corresponding values of the factor contained in the curly brackets 
in the integrand. The two further integrals that are required for the 
evaluation of $n(p)$ are calculated by a similar procedure. A more refined 
numerical method would be needed to obtain high statistical accuracy 
for a number of particles $N > 5$.

We report in the following our numerical results for $N = 2$. 
Figure 1 shows the function $\rho(x,0)$ as a function of $x$ for a 
pair of hard bosons, after suppression of the Gaussian decay imposed 
by the confinement through multiplication by $[\rho(x)\rho(0)]^{-1/2}$ 
where $\rho(x)$ is the particle density. 
The comparison with the same function for a pair of non-interacting 
fermions, for which a direct analytic calculation yields 
$\rho(x,0)/[\rho(x)\rho(0)]^{1/2} = (1 + 2x^2)^{-1/2}$, 
illustrates the role of statistics in determining the rate of decay 
of the density matrix. It may be remarked that the density matrix for 
the Bose gas reflects its spatial coherence and can be determined 
experimentally by measuring the interference pattern of two matter-wave 
beams emitted from two spatially separated regions of the trap~\cite{art17}.

Figure 2 reports the corresponding results that we obtain for the 
momentum distribution $n(p)$ in the case of two Bose or Fermi particles. 
The analytical result for fermions, with $n(p)$ described by the 
function $\pi^{-1/2}(1+2y^2)\exp(-y^2)$ where $y=p/(\hbar m\omega)^{1/2}$, 
is shown by the full line and serves to illustrate 
the accuracy of the Monte Carlo calculation.

Finally, Figure 3 tests the range of applicability of the asymptotic 
law given in Eq.~(\ref{eq7}) by reporting it in a log-log plot against the 
momentum distribution obtained in two Monte Carlo runs for a pair of 
hard bosons.

\section{Concluding remarks}

In this work we have demonstrated two aspects of the 
correlations which arise from Bose statistics in a 1D gas of hard, 
point-like bosons inside a harmonic trap.
A very remarkable feature of this system is the extremely slow decay of the
high-momentum tail in the single-particle momentum distribution, 
which we have determined analytically and verified numerically.
We have also shown that the one-body density matrix decays very slowly, 
after  elimination of the Gaussian factors that are imposed by the trap.

A consequence of the $p^{-4}$ tail in the momentum 
distribution is that the mean square fluctuation of the kinetic 
energy of the Tonks gas, which involves the fourth moment of $n(p)$, 
diverges under harmonic confinement. This property is even more 
remarkable if one considers that the kinetic energy density of the Tonks gas 
is the same as that of its fermionic map. This latter result, which can be 
demonstrated analytically in full detail from the bosonic and 
fermionic ground-state wave functions~\cite{art18}, may be viewed most 
directly to be a consequence of the basic theorem of density functional 
theory: the kinetic energy density is a unique functional of the 
particle density, which remains unchanged in the boson-fermion mapping.

We acknowledge support from INFM through PRA2001.

\begin{figure}
\centerline{\epsfig{file=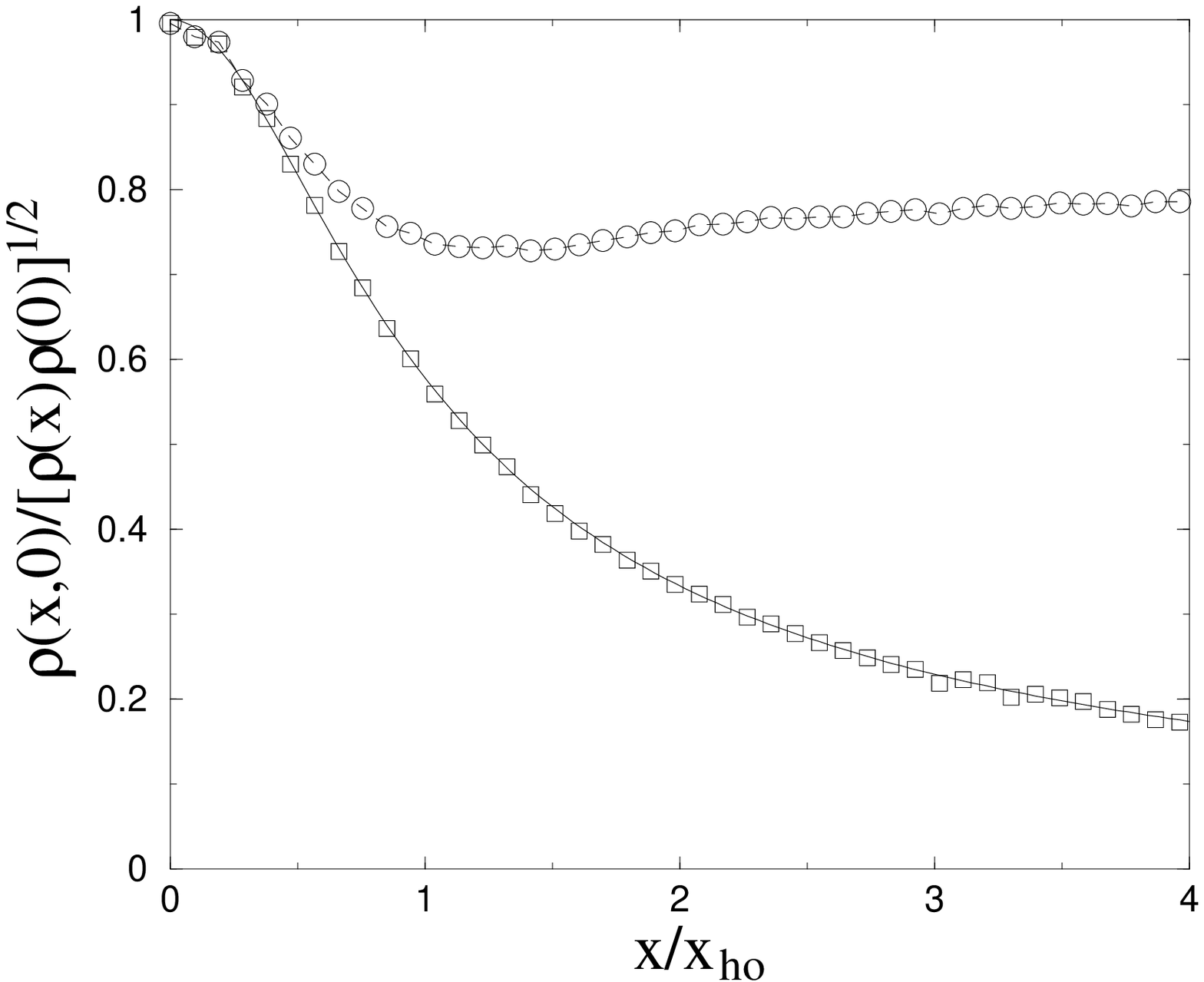,width=0.6\linewidth}}
\caption{Monte Carlo results for the one-body density 
matrix $\rho(x,x')/[\rho(x)\rho(x')]^{1/2}$, 
calculated as a function of $x$ at $x' = 0$, 
for a pair of hard point-like bosons (circles) and a pair of 
non-interacting fermions (squares) moving in 1D under harmonic 
confinement. The full line gives the exact analytical result for 
the case of fermions. The error bars in the Monte Carlo data are within
the size of the symbols.} 
\end{figure}

\begin{figure}
\centerline{\epsfig{file=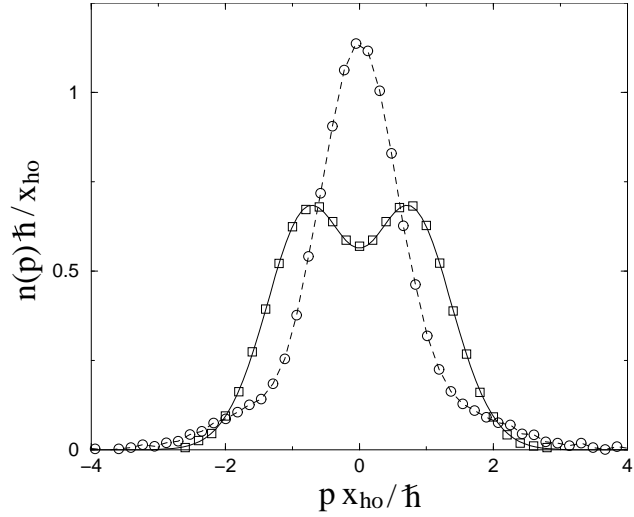,width=0.6\linewidth}}
\caption{Monte Carlo results for the momentum distribution of a pair 
of hard point-like bosons (circles) and a pair of non-interacting 
fermions (squares) moving in 1D under harmonic confinement. 
The full line gives the exact analytical result for the case of
fermions. The error bars in the Monte Carlo
data are within  the size of the symbols. 
}
\end{figure}

\begin{figure}
\centerline{\epsfig{file=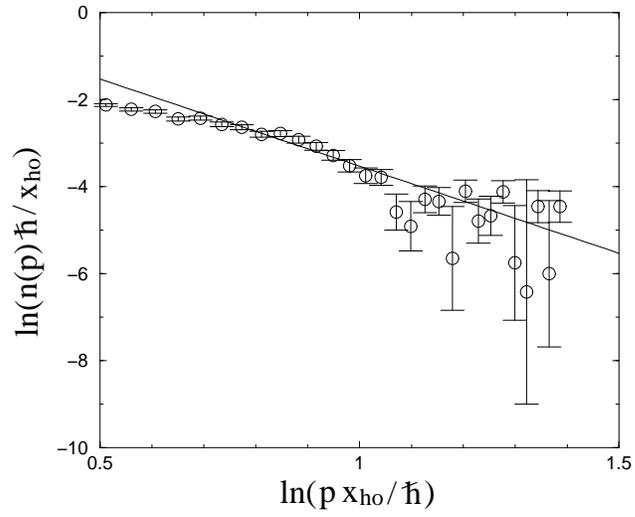,width=0.6\linewidth}}
\caption{Log-log plot of the momentum distribution of a pair of 
hard point-like bosons moving in 1D under harmonic confinement, 
from Monte Carlo integration (circles with error bars).
The full line shows 
the asymptotic law given in Eq. (\ref{eq7}). }
\end{figure}

\end{document}